\documentclass[aps,pra,reprint,groupedaddress]{revtex4-2}

\usepackage{graphicx}
\usepackage{amsmath}
\usepackage{hyperref}
\usepackage{booktabs}
\usepackage{longtable}

\begin{document}

\title{Time for unification: A single clock for Central and Eastern Europe}

\author{Chong Qi}
\email{chongq@kth.se}
\affiliation{Department of Physics, KTH Royal Institute of Technology, SE-10691 Stockholm, Sweden}

\date{\today}

\begin{abstract}
The EU's effort to abolish biannual clock changes has stalled. We argue this deadlock presents a historic opportunity for continental unification. Astronomical analysis demonstrates that strict alignment between clock and solar time is unnecessary; at high latitudes, extreme seasonal variations render the morning-versus-evening daylight debate physically irrelevant. Real-world examples demonstrate that societies adapt seamlessly to permanent solar offsets. We propose establishing a unified Central and Eastern European Time (CEET) at UTC+2. This converges naturally if Central Europe adopts permanent summer time while the Finnish-Baltic bloc retains permanent standard time, synchronizing the vast majority of member states under a single clock and eliminating cross-border economic friction.
\end{abstract}

\maketitle

\section{Introduction}

The practice of Daylight Saving Time (DST), or summer time, was originally conceived during the early 20th century by Germany as a wartime energy conservation measure. By shifting the clock forward one hour during the summer season, when the sun would otherwise rise at impractically early hours under standard time, policymakers aimed to use the early morning sunlight and extend the usable evening daylight, thereby reducing the consumption of coal and electricity for artificial illumination. A century later, modern energy grid analyses show that these savings are negligible, and in some cases, the increased use of air conditioning during longer summer evenings actually increases overall energy demand. At the same time, a substantial body of medical literature has documented acute health risks associated with the biannual clock transition, including short-term spikes in heart attacks, strokes, and traffic accidents caused by sleep disruption. A comprehensive 2017 Ex-post Impact Assessment by the European Parliamentary Research Service~\cite{eprs_2017} concluded that while summer time benefits the internal market and outdoor leisure, the biannual clock shifts cause significant circadian rhythm disruption and negative health outcomes.

In recent years, legislative efforts to abolish the biannual transition have gained momentum. In Europe, this shift was catalyzed by a 2018 public consultation by the European Commission~\cite{ec_survey_2018}, which revealed overwhelming support for abolishing the clock change across nearly all member states. The discontent is not merely statistical: in Finland, a citizens' petition to abolish the clock change in 2017 gathered more than 70,000 signatures, and in Sweden, a young schoolgirl famously wrote to then-Prime Minister Stefan L\"ofven urging him to end the practice, a letter that captured national attention and symbolized a generation's frustration with a policy it had never chosen. Following this public mandate, the European Parliament voted overwhelmingly in 2019 to end mandatory clock changes~\cite{eprs_2019}. However, this change was never implemented because the original EU proposal permitted each member state to independently choose whether to remain on permanent summer time or permanent winter time. This localized approach threatened administrative chaos and logistical complexity across highly integrated borders, requiring significant coordination efforts. The legislative effort has stalled ever since. In the United States, after years of debate surrounding the Sunshine Protection Act~\cite{us_sunshine_act}, the House of Representatives voted in 2026 to lock the nation into permanent DST; however, the bill has not yet cleared the Senate.

Recently, individual EU members have revived the debate. The debate has devolved into a deep divide between sleep scientists demanding permanent standard time (winter time) for circadian health, joined by education experts and parents who want morning sunlight for school children, on one side, and citizens, retail lobbies, and economists demanding permanent summer time on the other.
This divide is symptomatic of the paralyzed state of European Union institutions, which increasingly struggle to agree on any unified policy, a vulnerability that could prove fatal amid the current geopolitical turbulence and economic weakness facing the continent. 

This paper argues that solving the European deadlock offers an unprecedented opportunity: the bold consolidation of Central and Eastern Europe into a single, unified time zone, a necessary step to facilitate a truly frictionless single market and ensure long-term economic resilience.

\section{The socioeconomic and health dimensions of time}

The debate over time zones is frequently framed as a universal biological issue, but it is fundamentally dictated by astronomy, geometry, and latitude. The physical usefulness of DST is actually heavily concentrated in a specific mid-latitudinal ``sweet spot''. In this section, we examine the global context of time policies, the socioeconomic value of evening daylight, and the empirical health trade-offs of clock changes. To increase readability, the detailed technical analysis of the underlying astronomical mechanisms, including the Equation of Time, latitudinal daylight extremes, and detailed local solar tables, has been moved to the Appendices.

\subsection{The global situation}
More than 70 countries currently observe summer time~\cite{wiki_summer_time, dst_eu}, but the vast majority of these are located in Europe or other northern latitudes. For countries closer to the equator, where day length remains practically constant year-round, DST is not meaningful and is rarely observed. 

The United States sits significantly closer to the equator than northern Europe; only a small northern fraction of the country experiences the extreme winter morning darkness that makes early sunrises critical for safety. In principle, adopting seasonal clock shifts, or permanently advancing the clocks, makes little astronomical sense for the vast majority of the US population. Instead, the introduction of DST and its possible permanent adoption lean heavily on economic considerations. The push for extended evening daylight has been strongly supported by retail, tourism, and hospitality sectors, which benefit substantially from prolonged evening daylight that boosts the post-work outdoor economy. This exposes the ultimate reality of the time zone debate: modern time policies are often less about physiological health or solar alignment, and more about engineering an evening economy. This is not to dismiss the genuine benefits that evening light provides for outdoor activity and mental well-being, which we discuss below. As mentioned, the House of Representatives recently advanced legislation to adopt permanent summer time nationwide~\cite{us_sunshine_act}.

While China experimented with DST during 1940--1945 and again from 1986--1991, the practice was ultimately abandoned. Today, despite the single national time zone, over 90\% of China's population lives in the eastern coastal regions, placing them roughly in the ``correct'' solar time zone. In 2023, Egypt reintroduced DST with the explicit goal of reducing domestic electricity consumption. But independent analysts suggest that the actual impact was marginal. Starting in 1984, Russia aligned its clock-changing schedule with the rest of Europe. In 2011, Russia abolished DST and locked the entire country into permanent summer time. However, in 2014, it permanently shifted to standard (winter) time due to complaints of chronic fatigue, seasonal depression, and increased morning accidents.

The argument that a society will suffer extreme health crises if it does not strictly align with its geographic solar meridian is contradicted by real-world human adaptability. In 1940, Spain shifted its clocks ahead by an hour to align with Central Europe and never reverted. Despite being permanently out of sync with its solar time,  Spanish society  adapts by shifting its entire cultural schedule, and effectively maximizes their afternoon sun. France and the Benelux countries similarly operate on advanced clocks despite their western geography. Countries like Iceland and T\"{u}rkiye are in practice on permanent DST.

\subsection{The economics and psychology of the evening and the favor for summer time}

The advocates for permanent standard time argue that correct solar time is the only one compatible with human physiology. They warn that permanent summer time forces populations to start their days in biological darkness, resulting in chronic ``social jetlag''.
While this medical viewpoint focuses heavily on the morning commute, it largely dismisses the psychological and socioeconomic advantages of evening light that others prioritize. 

\subsubsection{Active vs. passive light and the psychological brake of afternoon darkness}
From an economic and societal perspective, one can argue that evening light is vastly more valuable than morning light. Morning light is ``passive.'' During the morning commute, people are focused primarily on commuting; extra sunlight at 07:30~AM does little to drive economic activity or social interaction, since most of that time is spent commuting, in an office, or in a classroom. 

Conversely, evening light is ``active.'' When people finish work or school, daylight directly translates to physical activity, outdoor sports, mental decompression, and consumer spending. Recent large-scale objective data confirms this behavioral shift; a 2026 causal analysis of wearable fitness tracker data~\cite{jeong_2026} found that while DST transitions produce no net change in total daily steps, they fundamentally reallocate physical activity: the ``spring forward'' transition significantly increases evening steps at the expense of morning activity, empirically validating the shift toward evening outdoor recreation. The history of the DST lobby in the United States also illustrates the economic value of this shift.

In northern latitudes, the psychological cost of standard time is profound. In Stockholm during the winter solstice, the sun sets around 14:50. This brutally early sunset acts as a psychological brake, effectively ending the day before working hours are over. Forcing the population into a bleak 3:00~PM sunset simply to achieve a slightly brighter 8:30~AM sunrise, which most people sleep through or ignore during their commute, is an unfavorable trade-off for mental health. Populations consistently express a strong preference for the extended evening daylight provided by summer time.

\subsubsection{The complex drivers of nightlife and evening activity}

\begin{figure}[htbp]
    \centering
    \includegraphics[width=\columnwidth]{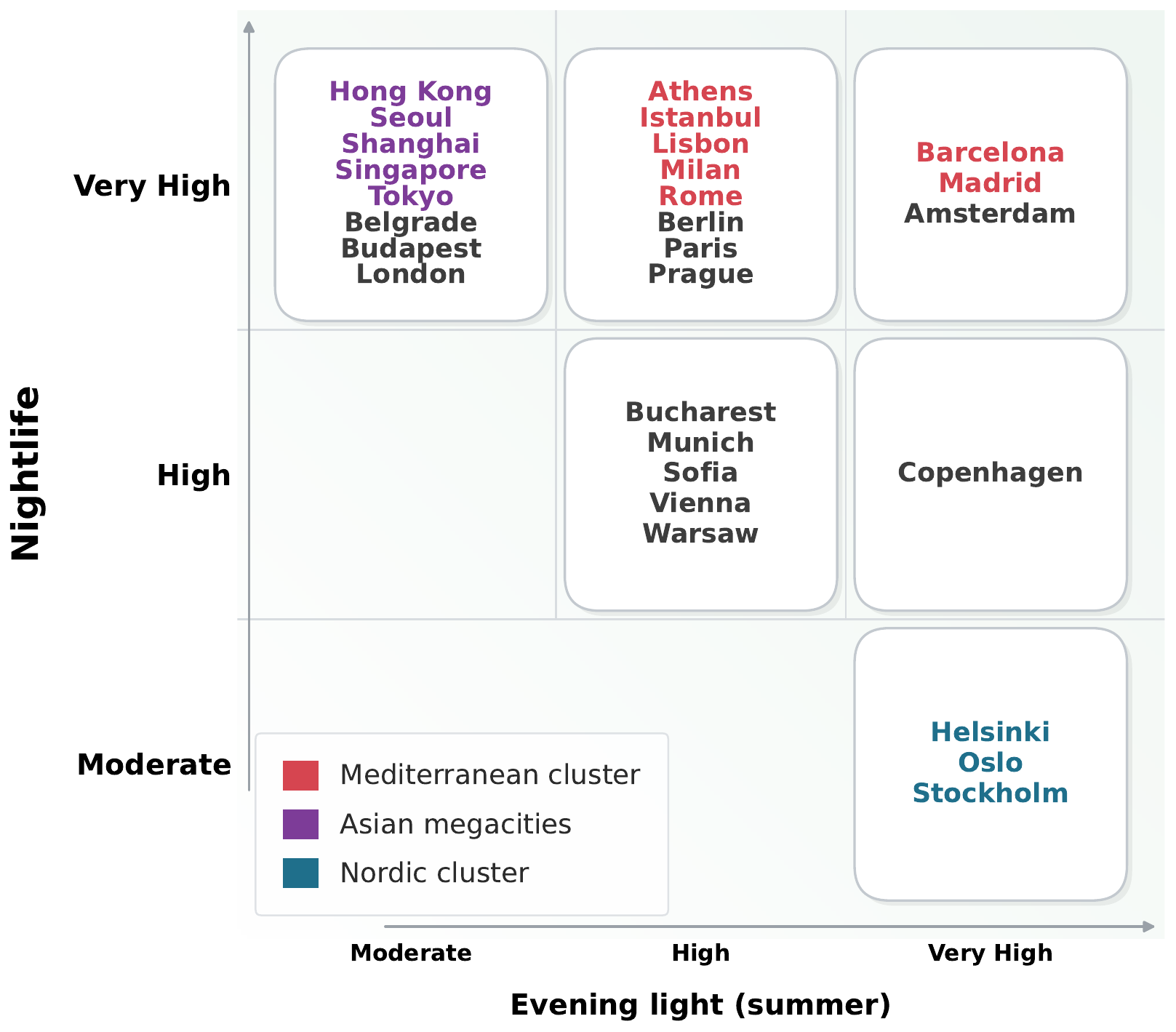}
    \caption{A comparative subjective classification of selected global cities matching summer evening light availability against cultural nightlife intensity. The matrix demonstrates that intense, late-night social cultures frequently develop and thrive even in regions with only moderate evening daylight.}
    \label{fig:nightlife}
\end{figure}

People generally believe that more evening daylight directly correlates with increased outdoor activity, particularly in the restaurant, sports, tourism, and shopping sectors. Yet, the relationship between evening light, time zone choice, and nightlife intensity is complex. As illustrated in Fig.~\ref{fig:nightlife}, daylight is only one factor among many---including population density, income levels, climate, and cultural norms---that shape evening economies.

For example, northern Europe enjoys an abundance of summer evening light but maintains a relatively subdued evening culture. Conversely, Asian metropolises like Hong Kong boast vibrant nightlife despite moderate evening light, driven largely by high urban density and culture. 

Late sunsets catalyze nightlife primarily when a society already has an entrenched habit of using the evening socially. The clock shift simply amplifies these existing traits. 
Spain and neighboring Portugal share almost identical longitude and solar time, but differ by exactly one hour of official clock time, providing a rare natural experiment. While Spain is arguably more famous for its late nightlife, there is no noticeable difference in vibrancy between the two countries. Recent research~\cite{bonmati2022wrongtime} shows that when corrected to a common solar reference, bedtime shows no significant difference between their populations. However, meal times remain significantly later in Spain, indicating that its late-hours culture is a shifted activity schedule rather than a shifted sleep schedule. 

While this misalignment between official and solar time carries a measurable physiological cost~\cite{bonmati2022wrongtime}, it should not be applied to national-level outcomes without qualification. Life expectancy and world happiness scores in Spain are actually higher than in Portugal~\cite{whr_2026}. This indicates that aggregate health outcomes are dominated by confounders (e.g., healthcare quality, diet) that eclipse any negative effect attributable to a one-hour clock offset. Time-zone choice must therefore be treated as just one behavioral variable within a much larger socioeconomic context.

\subsection{Health trade-offs: what the evidence actually shows}

The health dimension is frequently cited as the decisive argument against permanent summer time. However, the evidence is mixed. A major review~\cite{steponenaite_2026} found that while the spring transition correlates with short-term spikes in heart attacks and traffic accidents, the extra evening light reduces violent crime. In fact, the transition back to standard time each autumn is associated with a short-term decrease in all-cause mortality and workplace accidents, though the same review also finds that the autumn transition coincides with a rise in crimes involving physical harm.

While the spring transition undeniably causes temporary sleep loss and fatigue~\cite{owen_2022}, researchers describe the broader evidence as ``limited and heterogeneous.'' This challenges the medical community's assertion that permanent standard time is the only healthy choice. The medical community often focuses on population-level statistics—where a tiny increase in fatigue across millions causes a measurable spike in hospital visits—while ignoring that, for an individual, the transition is usually a minor annoyance. Sleep science identifies a real statistical cost but presents it as an individual health crisis. In the end, transition effects are small and transient, and the long-term health differences between permanent summer and winter time remain unclear.

If strict solar alignment were truly critical to human health, the world's longest-lived and happiest societies would cluster in their ``correct'' geographic time zones. The data show the opposite. Among the 25 most long-lived societies, only three sit in their correct geographic time zone year-round, while twenty are misaligned for at least half the year. A similar pattern emerges from the World Happiness Report~\cite{whr_2026}: only one of the top ten happiest countries sits in its correct geographic time zone year-round. Human well-being is determined by social infrastructure, healthcare, and community---not by the precise astronomical alignment of clock noon with solar noon. 
Iceland, despite enduring extreme winter darkness and living under permanent DST, consistently reports some of the lowest rates of seasonal affective disorder and highest rates of happiness globally. Modern populations have the technological and cultural tools to bypass astronomical light deficits, rendering the demand for strict solar alignment largely obsolete.

\section{Proposed scenarios for European reform}

If the European Union decides once again to abandon the biannual time change, the outcome will depend heavily on whether member states make a coordinated decision or act independently. Depending on the level of coordination, we anticipate the following possible scenarios:
\begin{itemize}
    \item \textbf{Scenario 1: Permanent standard or summer time.} Under a coordinated agreement, Europe could lock its existing time zones into either permanent summer time (CEST and EEST) or permanent standard time (CET and EET). While adopting permanent standard time feels more astronomically logical, as discussed above, there is a strong economic and social push for permanent summer time to preserve evening daylight. This seems the most logical choice but may require significant coordination efforts.

    \item \textbf{Scenario 2: Regional blocs.} If the EU simply removes the mandate and allows each country to decide on its own with no coordination, the result could be a varied patchwork of three time zones (a mix of CET, CEST/EET, and EEST). While this outcome grants maximum sovereignty and allows each nation to optimize for its specific solar and economic preferences, it would introduce new complexities for cross-border scheduling, transport, and the single market. But it is not totally chaotic: one might expect regional clusters like a Finnish-Baltic bloc and a Mediterranean bloc to maximize summer evening tourism. A new scenario may also emerge as explained below.

    \item \textbf{Scenario 3: The unified ``Central and Eastern European Time (CEET)''.} Rather than maintaining the current borders or allowing administrative fragmentation, removing the clock change could organically lead to a nearly unified European time zone through a few highly plausible steps:
    \begin{itemize}
        \item \textbf{Broad consensus to stop changing the clocks:} A strong majority of member states have already expressed a clear desire to abolish the mandatory spring and autumn clock changes entirely.
        \item \textbf{Central Europe prefers summer evenings:} Most CET countries are amenable to adopting permanent summer time (CEST) assuming their citizens value long, bright evenings for outdoor leisure.
        \item \textbf{Nordic countries may get brighter mornings or brighter evenings:} Finland may be concerned about extremely dark winter mornings and would likely prefer permanent standard time (EET) while other Nordic neighbors may accept CEST and brighter evenings. However, the Baltic states themselves favor permanent summer time and have been reluctant to diverge from Finland's choice, meaning Finland's decision would likely determine theirs by default~\cite{err2026baltic}.
        \item \textbf{The accidental merger (UTC+2):} Because Central Europe's permanent summer time (CEST) and the Finnish-Baltic bloc's permanent standard time (EET) are both exactly UTC+2, these two large blocks would organically merge. A large geographic swath of Europe would effortlessly harmonize on a single, unified time.
    \end{itemize}
\end{itemize}

The economic benefits of a single, synchronized European market can be immense. Traveling from Stockholm to Helsinki, or driving across the Polish-Lithuanian border, would no longer require a mental recalculation of the day. Missed international calls and confusing flight timetables would disappear. While this frictionless single market requires geographical ``compromises'', as Spain and France have already accepted, the friction of operating slightly out of sync with solar noon is ultimately outweighed by the frictionless trade, travel, and communication of a unified clock. Society would simply shift its opening hours to compensate.

\subsubsection{Economic and administrative costs of the status quo and future mixed time zone}
Beyond the cognitive burden, a decentralized system imposes measurable frictions on precisely the domains the internal market is designed to integrate:
\begin{itemize}
    \item Cross-border scheduling and labor mobility: workers and firms operating across the WET/CET/EET boundaries incur a persistent overhead in meeting coordination, shift planning, and contract timing that a single reference time would eliminate outright.
    \item Transport coordination: rail and aviation timetables that cross zone boundaries require explicit offset bookkeeping at every schedule change (including the twice-yearly DST transition), a recurring source of timetabling errors and missed connections.
    \item Grid and market operations: real-time electricity balancing and day-ahead market coupling across interconnected member states depend on unambiguous timestamping; a single EU time reference would remove an entire class of reconciliation error at zone and DST boundaries, a cost that compounds as cross-border renewable balancing becomes more time-critical.
\end{itemize}

\subsection{Why a single European time zone makes sense in a decentralized union}

Large countries and regions adopt different strategies for managing time across vast geographies, but they often function with much more coordination than their size suggests. In the United States, despite the existence of multiple time zones, Eastern Time acts as a de facto national reference because political, financial, and media centers are concentrated on the East Coast. Canada exhibits a similar, though weaker, pattern, with its largest population centers also located in the Eastern Time zone. Other geographically vast nations, such as Brazil, China, India, and Russia, are effectively organized around one dominant national time, regardless of how many official time zones they formally maintain.

The European Union, however, differs fundamentally from these centralized examples. Economic, political, and cultural functions are deeply decentralized, distributed across multiple cities, 27 sovereign member states, and associated neighboring countries, rather than concentrated in a single dominant center. As a result, there is no universally recognized European reference time comparable to Eastern Time, even though CET is arguably dominant to some extent.

Because of this profound decentralization, the EU's current patchwork of three official time zones (WET, CET, EET) is uniquely burdensome. In everyday life, Europeans rarely refer to formal time-zone designations. Instead, people think in terms of national borders, speaking of ``Swedish time,'' ``German time,'' or ``Finnish time.'' While many Europeans know the time zones of their immediate neighbors, edge cases cause frequent confusion. For example, many Swedes know that Finland is one hour ahead, but few immediately recognize that Mariehamn in the {\AA}land Islands follows Finnish time despite being geographically adjacent to Sweden. The straight-line distance between Stockholm and Helsinki is only about 400~km, corresponding to a true solar time difference of just 27 minutes.

This reliance on national borders to dictate time creates a significant cognitive burden for the Union. The challenge is further amplified by the borderless nature of the Schengen Area. A traveler or cross-border commuter may cross a time-zone boundary while moving only a short physical distance without any visible checkpoints. Because modern electronic devices automatically and silently update local time, individuals frequently become unaware that the clock has changed, leading to a high risk of missed meetings, transport delays, and scheduling failures.

This leads to a stark mathematical reality regarding coordination:
\begin{itemize}
    \item \textbf{A centralized, single-zone system strictly requires navigating just one time zone.}
    \item \textbf{A system with even two time zones effectively requires navigating 27 individual national clocks.} While geographic rules exist, tracking these invisible boundaries without a map is highly impractical for visitors or anyone unfamiliar with European geography. 
\end{itemize}

By adopting a single European time zone, the EU would resolve the inherent friction of its decentralized political structure, instantly eliminating the cognitive overhead required to navigate 27 independent national clocks.

\begin{figure*}[htbp]
    \centering
    \includegraphics[width=\textwidth]{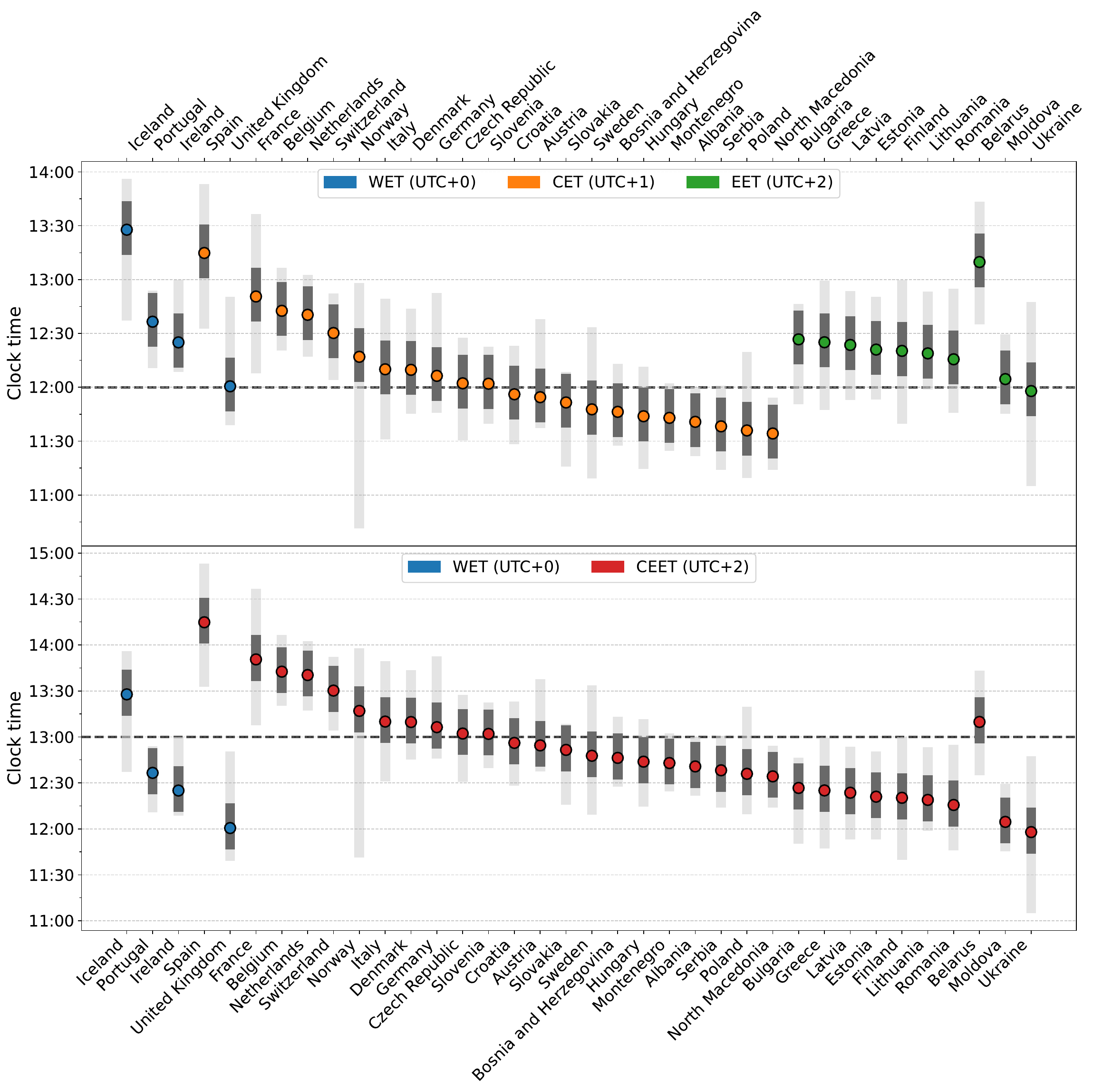}
    \caption{Mean solar noon and total geographical clock displacement across Europe. Circles indicate the clock time of mean solar noon at each nation's capital. Dark solid bars show seasonal fluctuations caused by the Equation of Time (EoT)~\cite{wiki_eot, meeus_1998}, while light grey bars illustrate the total standard time variance across the contiguous country's longitudinal span (excluding overseas and remote territories). Top Panel: Current standard time configuration (WET, CET, EET). Bottom Panel: A simulated unified scenario where CET nations adopt permanent summer time, merging CET and EET into a single ``CEET'' block at UTC+2.}
    \label{fig:solar_noon_scenarios}
\end{figure*}

\begin{figure*}[htbp]
    \centering
    \includegraphics[width=\textwidth]{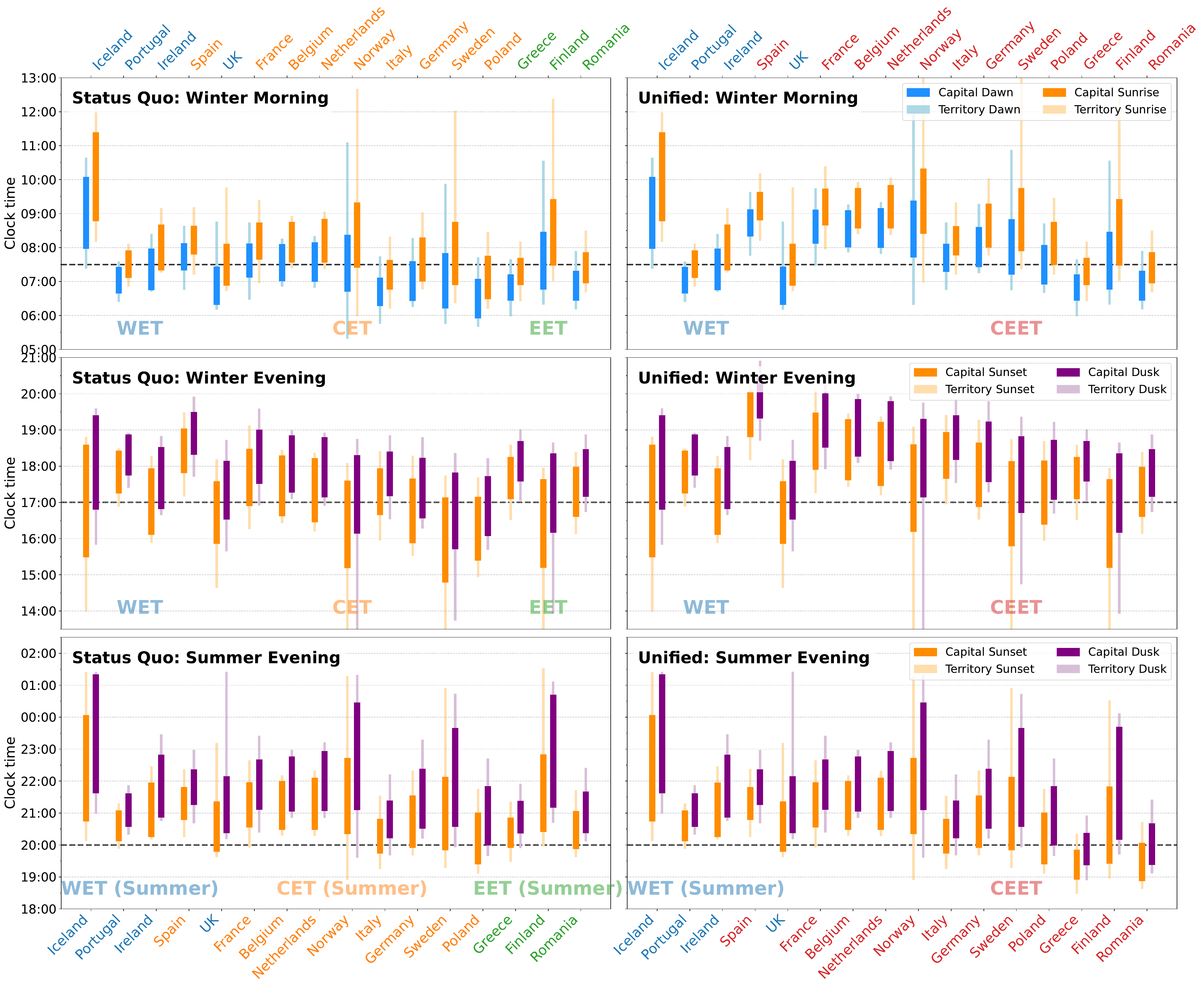}
    \caption{Extreme seasonal ranges of dawn, sunrise, sunset, and dusk across selected contiguous European nations. Left column: Current standard time (Status Quo). Right column: Simulated unified scenario with a permanent UTC+2 CEET block (WET unchanged). Top Panels: Winter Morning. Middle Panels: Winter Evening. Bottom Panels: Summer Evening. Thin central bars denote the natural variation of solar events at each nation's capital due to solar declination and the EoT. Thicker, lighter background bars represent the absolute temporal extremes across the nation's entire contiguous territory. For Nordic summer evenings, upper bounds reflect the last observed sunset/dusk before Midnight Sun. Remote and overseas territories are excluded to prevent distortion.}
    \label{fig:daylight_extremes}
\end{figure*}

\subsection{Consequences and anticipated objections}
The most common objection is that a single time zone would create a large mismatch between clock time and solar time at the geographic extremes (e.g., very late sunrises in the west, very early sunrises in the east). 
As shown in Fig.~\ref{fig:solar_noon_scenarios}, under this unified CEET clock, solar noon forms a continuous, rising geographical gradient across the continent. Ten capitals will experience solar noon later than 13:00 in winter (which they already experience in summer).
But it is a fixed, predictable cost borne by geography, not a variable, unpredictable cost borne by political negotiation. Individuals and institutions can adapt permanently to a known solar offset. 

Under the unified CEET scenario, all CET countries will be able to enjoy some period of after-work light during the winter (see Fig.~\ref{fig:daylight_extremes}). However, as a direct consequence, they will have to endure notably darker mornings. 

The current European time-zone arrangement has an interesting and elegant feature: despite large differences in latitude, many countries using CET have surprisingly similar clock times for sunrise in winter. This is a result of a balance between geography and the artificial time-zone system (see Fig.~\ref{fig:daylight_extremes} and Appendix~\ref{sec:appendix_latitudinal_extremes}). If CET countries were to adopt permanent summer time, this balance would shift: all of them would experience later winter sunrises, with several months of mornings becoming significantly darker by the clock. In northern Europe especially, the period with sunrise after 08:00 would roughly double from around two months to four months. This effect cannot be avoided unless clocks are shifted in the opposite direction by another hour; therefore, the debate comes down to a choice between darker mornings and darker evenings.

In this proposal, the main speculated economic pressure would be felt by southern and eastern European countries such as Romania, Greece, Cyprus, and Bulgaria. Because they rely heavily on their Mediterranean evening cultures and summer tourism, they may fear that reverting to UTC+2 would hurt their nightlife businesses. It is likely that they would opt for permanent summer time (EEST). Nevertheless, one can still expect some sunshine or twilight until 20:00, much later than in Asian megacities like Hong Kong. 

\section{Summary and discussion}

The clock-change debate often appears to be yet another deadlocked European policy, stalled indefinitely between broad public consensus and a lack of governmental coordination. Yet this deadlock presents a historic opportunity: establishing a unified time zone across most of the continent.

This paper argues that the standoff can be resolved by shifting the objective from optimizing local solar time to prioritizing continental coordination. Astronomical analysis shows that strict alignment between clock time and solar time is neither achievable nor necessary. High-latitude seasonal extremes render the morning-versus-evening debate largely irrelevant, while populations in Spain, France, and Iceland demonstrate that societies adapt effortlessly to permanent solar offsets with no measurable harm to health or happiness.

The idea of a ``correct'' time zone is fragile. Driven by the Equation of Time (EoT)~\cite{wiki_eot, meeus_1998}, natural shifts in solar noon mean that only four EU states maintain solar alignment year-round---and even then, only during winter time. The vast majority of EU citizens already live under a clock that does not match their true solar time. This quiet, continent-wide reality is a powerful argument against the pursuit of solar perfection and in favor of pragmatic unification.

The proposed unified time zone relies on a fortunate coincidence: Central Europe's preferred permanent summer time and the Finnish-Baltic bloc's preferred permanent standard time are exactly the same clock. While Norway and Sweden might initially prefer permanent winter time to avoid darker mornings, the benefits of coordinating with Germany and enjoying lighter winter evenings would likely outweigh these concerns. By recognizing this natural convergence, the EU can unite the vast majority of its territory under a single time reference. This would eliminate the cognitive burden of 27 national clocks, reduce friction in cross-border markets, and end the twice-yearly disruption. The cost is merely a predictable solar offset that a modern country can successfully manage today. 

At the geographical extremes, minor deviations would remain but still align organically. In the southeast, nations like Romania, Greece, Cyprus, and Bulgaria might opt for an even more advanced clock to preserve their celebrated nightlife economies, naturally aligning with neighboring T\"{u}rkiye. On the western flank, Portugal, Ireland, and the United Kingdom could favor summer time to prevent a severe two-hour gap with the unified European bloc. Iceland already operates on what is effectively permanent summer time, functioning well despite an extreme solar offset. Taken together, these plausible national preferences suggest that an organic convergence toward a largely unified European clock is much closer than the current political deadlock implies.

\section*{Acknowledgments}
This work was partly inspired by a news article~\cite{err2026baltic}. All data presented in this work were collected with the help of Google Gemini.

\appendix

\section{Astronomical theory of noon and the Equation of Time}
\label{sec:appendix_astronomical_theory}
Historically, societies used true solar noon, the moment the sun reaches its highest point. However, this timing is irregular due to Earth's orbital eccentricity and axial tilt. The difference between true solar time and a constant clock is the Equation of Time (EoT), approximated in minutes by~\cite{wiki_eot, meeus_1998}:
\begin{equation}
\text{EoT}(n) \approx -7.65 \sin\left(\frac{2\pi}{365}(n - 2)\right) + 9.87 \sin\left(\frac{4\pi}{365}(n - 80)\right)
\end{equation}
where $n$ is the day of the year. While the EoT is relatively small in summer, it causes notable variations in autumn and winter, shifting solar noon by up to $\pm 15$ minutes. Modern timekeeping uses mean solar noon to resolve this, creating a steady 24-hour day. The advent of railways further required standard time zones. The relationship between true solar time ($T_{\text{solar}}$) and local standard clock time ($T_{\text{clock}}$) is:
\begin{equation}
T_{\text{solar}} = T_{\text{clock}} + 4 \times (L_{\text{local}} - L_{\text{meridian}}) + \text{EoT}
\end{equation}
where $L_{\text{local}}$ and $L_{\text{meridian}}$ are the local and reference longitudes (in degrees east).

Because Europe stretches across vast longitudes under few time zones, clock time already deviates significantly from the sun (see Table~\ref{tab:solar_noon}). While Prague and Berlin align well under standard time, Paris and Madrid operate hours ahead of their solar time. During summer time (DST), every city misses astronomical noon. For instance, solar noon in summer-time Madrid occurs at 14:20, a considerable shift that enables its late-night culture and suggests that populations prioritize social evening light over astronomical midday.

\section{Latitudinal extremes, twilight, and social time}
\label{sec:appendix_latitudinal_extremes}
Unlike noon, sunrise and twilight depend heavily on latitude. They are derived from the solar hour angle $\omega$:
\begin{equation}
\cos \omega = \frac{\sin h_0 - \sin \phi \sin \delta}{\cos \phi \cos \delta}
\end{equation}
where $\phi$ is latitude, $\delta$ is solar declination, and $h_0$ is the solar elevation angle ($-0.833^\circ$ for sunrise/sunset, $-6^\circ$ for civil twilight). The event times are symmetric around noon:
\begin{align}
T_{\text{sunrise/dawn}} &= T_{\text{noon}} - \frac{\omega}{15^\circ} \\
T_{\text{sunset/dusk}} &= T_{\text{noon}} + \frac{\omega}{15^\circ}
\end{align}

These equations demonstrate that, theoretically, the traditional logic of DST, shifting daylight from early mornings to evenings, only works optimally between $46^\circ$N and $54^\circ$N. Outside this band, the logic breaks down. In northern Europe, seasonal daylight swings are extreme. In winter, the sun rises late and sets as early as 15:00. This premature afternoon darkness is psychologically taxing. While morning light is desirable, prioritizing it at high latitudes is astronomically impossible without sacrificing the entire afternoon. For example, ensuring a 07:00 winter sunrise in Stockholm would require an extreme clock shift to UTC-1, forcing sunset at an unworkable 12:45. 

Forcing northern countries into permanent standard time simply to save an hour of morning light achieves very little while maximizing evening darkness. In these extreme latitudes, what matters far more is the prolonged duration of northern twilight, which provides a gradual waking signal that naturally compensates for the lack of direct morning sunlight.

This latitudinal dependence also explains an intriguing feature of the European winter: why a northern city like Stockholm and a southwestern city like Madrid can experience sunrise at roughly the same clock time. The locus of places experiencing sunrise at the same instant---the equal-sunrise line, or terminator---is determined by both longitude and latitude. While longitude alone determines local solar noon, latitude dictates how the sun's path intersects the horizon. From spherical astronomy, setting the solar elevation to zero ($h_0 = 0$) yields the sunrise hour angle $\omega_s$:
\begin{equation}
 \omega_s = \arccos(-\tan \phi \tan \delta)
\end{equation}
During winter, the sunrise curve tilts sharply, cutting from the southwest (Spain) toward the northeast (Sweden). As a result, these vastly separated regions intersect the terminator curve simultaneously, sharing remarkably similar winter sunrise times.

\onecolumngrid
\section{Local solar and twilight times across Europe}

\begin{longtable}{lcccccccc}
\caption{Local clock times of solar noon and civil twilight (dawn and dusk) across European capital cities on the winter solstice (Dec 21) and summer solstice (Jun 21). Times are formatted to the respective local winter and summer time zones. The duration column displays the total twilight duration (dusk minus dawn), with the actual sunshine duration (sunset minus sunrise) shown in brackets. (*) Reykjavik and Minsk maintain a permanent year-round time zone.}
\label{tab:solar_noon} \\
\toprule
\textbf{City} & \textbf{Winter} & \textbf{Winter} & \textbf{Winter} & \textbf{Winter Dur.} & \textbf{Summer} & \textbf{Summer} & \textbf{Summer} & \textbf{Summer Dur.} \\
& \textbf{Noon} & \textbf{Dawn} & \textbf{Dusk (PM)} & \textbf{Twi [Sun]} & \textbf{Noon} & \textbf{Dawn} & \textbf{Dusk (PM)} & \textbf{Twi [Sun]} \\
\midrule
\endfirsthead

\multicolumn{9}{c}%
{{\bfseries \tablename\ \thetable{} -- continued from previous page}} \\
\toprule
\textbf{City} & \textbf{Winter} & \textbf{Winter} & \textbf{Winter} & \textbf{Winter Dur.} & \textbf{Summer} & \textbf{Summer} & \textbf{Summer} & \textbf{Summer Dur.} \\
& \textbf{Noon} & \textbf{Dawn} & \textbf{Dusk (PM)} & \textbf{Twi [Sun]} & \textbf{Noon} & \textbf{Dawn} & \textbf{Dusk (PM)} & \textbf{Twi [Sun]} \\
\midrule
\endhead

\midrule \multicolumn{9}{r}{{Continued on next page}} \\
\endfoot

\bottomrule
\endlastfoot
\midrule
\multicolumn{9}{c}{\textbf{WET/GMT}} \\
\midrule
London & 11:58 & 7:23 & 4:34 & 9h11m [7h48m] & 1:02 & 3:54 & 10:09 & 18h15m [16h37m] \\
Dublin & 12:22 & 7:54 & 4:51 & 8h56m [7h29m] & 1:26 & 4:03 & 10:50 & 18h46m [16h59m] \\
Lisbon & 12:34 & 7:20 & 5:48 & 10h27m [9h26m] & 1:38 & 5:39 & 9:37 & 15h57m [14h52m] \\
Reykjavik* & 1:25 & 10:02 & 4:49 & 6h47m [4h05m] & 1:29 & -- & -- & 24h00m [21h07m] \\
\midrule
\multicolumn{9}{c}{\textbf{CET}} \\
\midrule
Madrid & 1:12 & 8:03 & 6:22 & 10h19m [9h16m] & 2:16 & 6:11 & 10:22 & 16h10m [15h03m] \\
Paris & 12:48 & 8:03 & 5:33 & 9h30m [8h14m] & 1:52 & 5:03 & 10:41 & 17h37m [16h10m] \\
Brussels & 12:40 & 8:02 & 5:18 & 9h16m [7h55m] & 1:44 & 4:41 & 10:46 & 18h05m [16h30m] \\
Luxembourg & 12:33 & 7:51 & 5:16 & 9h24m [8h07m] & 1:37 & 4:43 & 10:31 & 17h47m [16h17m] \\
Amsterdam & 12:38 & 8:06 & 5:10 & 9h04m [7h39m] & 1:42 & 4:27 & 10:56 & 18h29m [16h47m] \\
Berlin & 12:04 & 7:32 & 4:36 & 9h03m [7h38m] & 1:08 & 3:52 & 10:24 & 18h31m [16h49m] \\
Stockholm & 11:45 & 7:47 & 3:44 & 7h57m [6h03m] & 12:49 & 1:57 & 11:41 & 21h44m [18h36m] \\
Oslo & 12:14 & 8:20 & 4:10 & 7h50m [5h52m] & 1:18 & 2:07 & 12:30 & 22h23m [18h48m] \\
Copenhagen & 12:07 & 7:49 & 4:25 & 8h35m [7h00m] & 1:11 & 3:23 & 10:59 & 19h36m [17h31m] \\
Rome & 12:07 & 7:02 & 5:14 & 10h11m [9h07m] & 1:11 & 5:00 & 9:23 & 16h23m [15h13m] \\
Vienna & 11:52 & 7:05 & 4:39 & 9h34m [8h19m] & 12:56 & 4:11 & 9:40 & 17h29m [16h04m] \\
Prague & 12:00 & 7:19 & 4:41 & 9h21m [8h02m] & 1:04 & 4:07 & 10:00 & 17h53m [16h22m] \\
Budapest & 11:41 & 6:52 & 4:31 & 9h39m [8h25m] & 12:45 & 4:05 & 9:25 & 17h20m [15h57m] \\
Warsaw & 11:33 & 7:01 & 4:06 & 9h05m [7h41m] & 12:37 & 3:24 & 9:51 & 18h26m [16h46m] \\
Bratislava & 11:49 & 7:02 & 4:37 & 9h34m [8h20m] & 12:53 & 4:09 & 9:37 & 17h28m [16h03m] \\
Ljubljana & 11:59 & 7:06 & 4:54 & 9h48m [8h37m] & 1:03 & 4:31 & 9:35 & 17h04m [15h45m] \\
Zagreb & 11:54 & 6:59 & 4:49 & 9h49m [8h39m] & 12:57 & 4:27 & 9:28 & 17h01m [15h43m] \\
Bern & 12:28 & 7:37 & 5:19 & 9h42m [8h30m] & 1:31 & 4:55 & 10:09 & 17h14m [15h52m] \\
Belgrade & 11:36 & 6:38 & 4:34 & 9h55m [8h46m] & 12:39 & 4:14 & 9:05 & 16h50m [15h34m] \\
Sarajevo & 11:44 & 6:44 & 4:45 & 10h01m [8h53m] & 12:48 & 4:27 & 9:08 & 16h41m [15h27m] \\
Podgorica & 11:40 & 6:36 & 4:45 & 10h08m [9h03m] & 12:44 & 4:30 & 8:58 & 16h28m [15h17m] \\
Tirana & 11:38 & 6:31 & 4:46 & 10h14m [9h10m] & 12:42 & 4:33 & 8:51 & 16h18m [15h09m] \\
Skopje & 11:32 & 6:26 & 4:38 & 10h11m [9h06m] & 12:36 & 4:23 & 8:48 & 16h24m [15h14m] \\
Valletta & 11:59 & 6:39 & 5:20 & 10h41m [9h42m] & 1:03 & 5:15 & 8:52 & 15h37m [14h35m] \\
\midrule
\multicolumn{9}{c}{\textbf{EET}} \\
\midrule
Helsinki & 12:18 & 8:24 & 4:11 & 7h46m [5h47m] & 1:22 & 1:58 & 12:45 & 22h47m [18h54m] \\
Tallinn & 12:18 & 8:21 & 4:17 & 7h55m [6h01m] & 1:22 & 2:27 & 12:17 & 21h51m [18h38m] \\
Riga & 12:21 & 8:10 & 4:33 & 8h23m [6h42m] & 1:25 & 3:20 & 11:30 & 20h10m [17h51m] \\
Vilnius & 12:16 & 7:54 & 4:39 & 8h44m [7h13m] & 1:20 & 3:43 & 10:57 & 19h13m [17h17m] \\
Bucharest & 12:13 & 7:14 & 5:12 & 9h57m [8h49m] & 1:17 & 4:53 & 9:40 & 16h47m [15h32m] \\
Sofia & 12:24 & 7:21 & 5:28 & 10h07m [9h01m] & 1:28 & 5:13 & 9:43 & 16h30m [15h19m] \\
Athens & 12:23 & 7:07 & 5:38 & 10h31m [9h31m] & 1:26 & 5:30 & 9:22 & 15h51m [14h47m] \\
Nicosia & 11:44 & 6:22 & 5:06 & 10h44m [9h46m] & 12:48 & 5:02 & 8:34 & 15h32m [14h31m] \\
Kyiv & 11:55 & 7:16 & 4:35 & 9h19m [7h59m] & 12:59 & 4:00 & 9:59 & 17h59m [16h26m] \\
Chisinau & 12:02 & 7:11 & 4:53 & 9h42m [8h29m] & 1:06 & 4:29 & 9:43 & 17h14m [15h53m] \\
Minsk* & 1:07 & 8:41 & 5:33 & 8h51m [7h22m] & 1:11 & 3:42 & 10:40 & 18h57m [17h06m] \\
\end{longtable}

\twocolumngrid

\end{document}